\documentclass{aa}  

\usepackage{graphicx}
\usepackage{txfonts}
\begin{document}

\title{An episodically variable stellar wind in the planetary nebula IC\,4997\thanks{Based on observations collected at the Centro Astron\'omico
    Hispano-Alem\'an (CAHA) at Calar Alto, operated jointly by Junta de Andaluc\'{\i}a and Consejo Superior de Investigaciones Cient\'{\i}ficas (IAA-CSIC).}}

   \author{Luis F. Miranda\inst{1}
          \and
          Jos\'e M. Torrelles\inst{2,3}
          \and
          Jorge Lillo-Box\inst{4}
          }

   \institute{Instituto de Astrof\'{\i}sica de Andaluc\'{\i}a--CSIC, Glorieta de la
     Astronom\'{\i}a s/n, 18008, Granada, Spain\\
     \email{lfm@iaa.es}
     \and
     Institut de Ci\`encies de l'Espai (ICE, CSIC), Can Magrans s/n, E-08193 Cerdanyola del Vall\`es, Barcelona, Spain
     \and
     Institut d'Estudis Espacials de Catalunya (IEEC), Barcelona, Spain
     \and
         Departamento de Astrof\'{\i}sica, Centro de Astrobiolog\'{\i}a (INTA-CSIC), ESAC
         Campus, Camino Bajo del Castillo s/n, 28692, Villanueva de la Ca\~nada, Madrid, Spain\\
             \email{jlillo@cab.inta-csic.es}
             }

   \date{Received ; accepted }

 
   \abstract{IC\,4997 is a planetary nebula well known by its variability. We present
     high-resolution spectra of IC\,4997 obtained in 1993, 2019, and 2020 that reveal changes
     in the H$\alpha$ and [N\,{\sc ii}] emission line profiles, which had never been reported for this object.
     The H$\alpha$ P\,Cygni emission profile observed in 1993 changed to a single-peaked
     profile in 2019--2020, implying that the stellar wind has largely weakened. The very broad H$\alpha$ emission wings
     narrowed by a factor of $\sim$2 between 1993 and 2019--2020, indicating that the efficiency of the Rayleigh--Raman
     scattering has noticeably decreased. A high-velocity [N\,{\sc ii}] nebular component detected in 1993 is missing
     in 2019 and 2020, probably due to a decrease in its electron density. A correlation exists between the strength
     of the stellar wind
     and the episodic ($\sim$50--60\,yr) variation in the [O\,{\sc iii}]$\lambda$4363/H$\gamma$ line
     intensity ratio, suggesting that
     an episodic, smoothly variable stellar wind is the main cause of the variability of IC\,4997.
     Monitoring of that intensity ratio and of the H$\alpha$ emission line profile in the coming years and new
     multiwavelength observations 
     are key to unveiling the ongoing processes in IC\,4997 and constraining the origin of the wind variability.}

   \keywords{planetary nebulae: individual: IC\,4997 -- circumstellar matter --
     stars: winds and outflows -- ISM: jets and outflows }

   \titlerunning{Variable stellar wind in the
     planetary nebula IC\,4997}

   \maketitle
%

\section{Introduction}

   After leaving the asymptotic giant branch (AGB), a star evolves toward higher effective 
   temperatures ($T$$_{\rm eff}$). When $T$$_{\rm eff}$ reaches $\sim$25000\,K, the radiation of the
   central star (CS) photoionizes the envelope expelled during the AGB, forming a
   planetary nebula (PN). The natural variation in $T$$_{\rm eff}$ and the expansion of the
   ionized envelope result in changes of the nebular parameters and emission line intensities.
   Thus, all PNe are expected to be variable at a rate dictated by the evolutionary timescales
   of the CS and of the nebular expansion. This secular variability has been observed in
   several PNe (Hajduk et al. 2015). On the other hand, some PNe present a
   variability that differs from the secular one. These PNe are particularly interesting
   because they reveal the existence of processes in the PN phase that deviate from the normal
   evolution of CSs (e.g., Feibelman et al. 1992; Hyung et al. 1994; Reindl et al. 2017).

   Among variable PNe, \object{IC\,4997} has been considered a paradigmatic case since Liller \& Aller (1957) and
   Aller \& Liller (1966) discovered variability in its [O\,{\sc iii}]$\lambda$4363/H$\gamma$ 
   line intensity ratio. Since then, multiple observations have shown that this ratio could 
   vary with a period of $\sim$50--60\,yr and that the V magnitude of the nebula, additional nebular emission
   lines and emission line ratios, the electron temperature ($T$$_{\rm e}$) and electron density ($N$$_{\rm e}$),
   and the radio continuum flux density of the nebula and of particular nebular regions are also variable
   (Miranda \& Torrelles 1998, hereafter MT98; Kostyakova \& Arkhipova 2009; Burlak \& Esipov
   2010; Hajduk et al. 2018; Arkhipova et al. 2020, hereafter A+20). The origin of the variability has not been
   unraveled: an outburst, wind variability, nebular expansion, changes in $T$$_{\rm eff}$, $T$$_{\rm e}$,
   and/or $N$$_{\rm e}$ have all been suggested, among other possibilities (MT98; A+20 and references therein).

\begin{figure*}
  \centering
        \includegraphics[width=18.0cm]{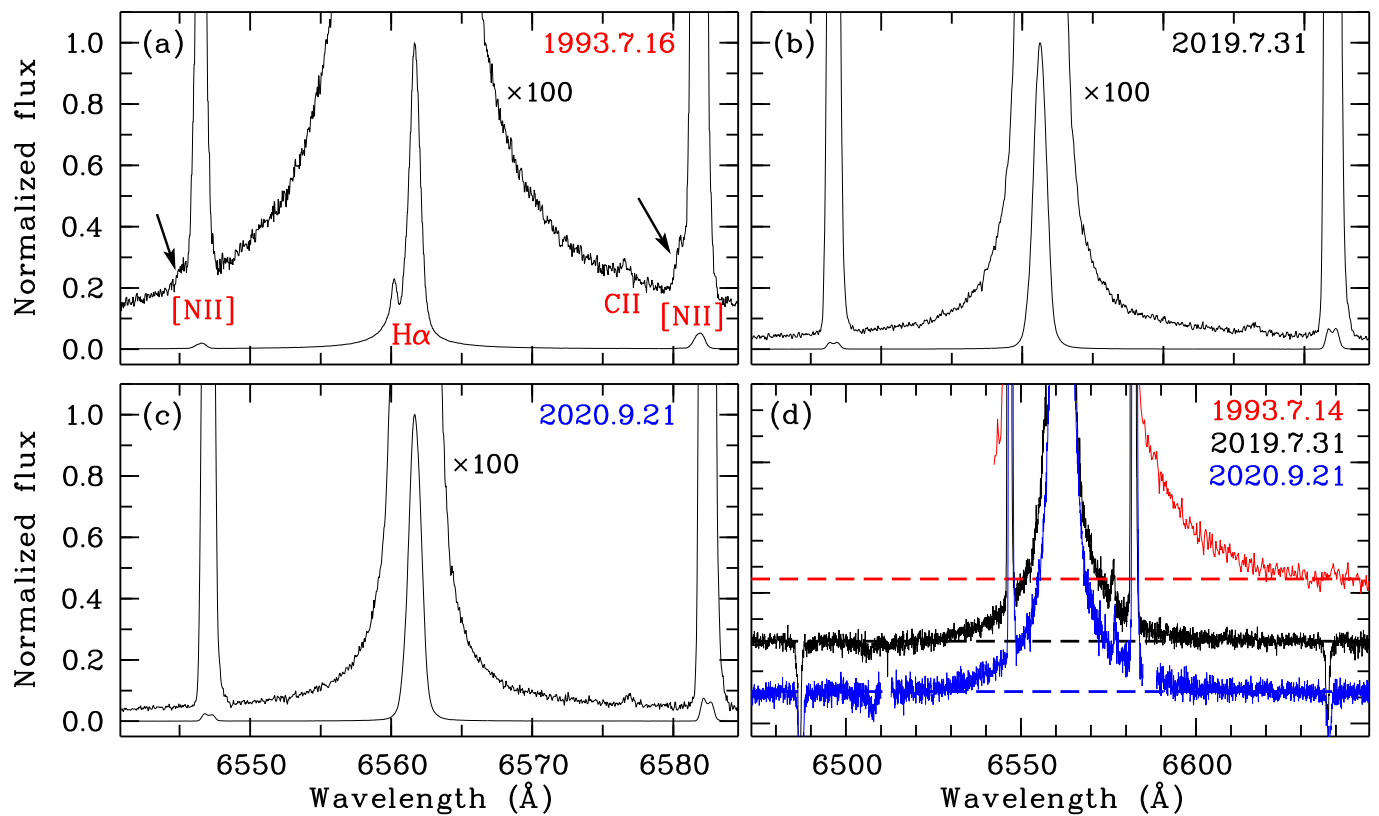}
        \caption{Profiles of the H$\alpha$ emission line. {\it (a), (b), (c)}.
          High-resolution spectra of IC\,4997 between 6450 and
          6670\,{\AA} at the three epochs. The spectra are normalized
          to the intensity peak of the H$\alpha$ emission line in each epoch and are shown at
          two different scales. Panel {\it (a)} marks the observed emission lines and
          the [N\,{\sc ii}] HVC detected in 1993 (see also Fig.\,3).
          {\it (d)} Spectra at the three epochs between 6473 and 6649.5\,{\AA,} 
          arbitrarily shifted vertically. The 1993 spectrum is adapted from MTE96 and covers
          from 6542\,{\AA}. The horizontal line associated with each spectrum represents the stellar
          continuum extrapolated from line-free regions. The absorption at $\sim$6637.8\,{\AA} in the
            2019 and 2020 spectra {\it (d)} does not correspond to the O\,{\sc i}\,$\lambda$1027\,{\AA} Raman feature identified in
            Orion (Henney 2021) and could be telluric.}
\end{figure*}

IC\,4997 also has other interesting characteristics. It is a very young PN, with a kinematical
age of $\sim$670\,yr (if located at 2.5\,kpc), that consists of an outer bipolar and an inner, very
compact elliptical shell, as first revealed by radio continuum observations at 3.6 and 2\,cm wavelengths
(Miranda et al. 1996, hereafter MTE96; MT98). Evidence for a dense equatorial ring-like structure has been
found at the 2\,cm radio continuum and in the 3.6--2\,cm spectral index map (MT98). The H$\alpha$ emission line presents
very broad wings ($>$5400\,km\,s$^{-1}$; MTE96) that are explained by Rayleigh-Raman scattering
in the H\,I neutral envelope around the object (Altschuler et al. 1986; Lee \& Hyung 2000,
hereafter LH00). The dust content of the nebula is very high (Pottasch et al. 1984), and several
molecules have been detected in it (Rao et al. 2020 and references therein). The CS is classified
as a {\it wels} or [WC10] type, and its $T$$_{\rm eff}$ is $\sim$55000 (see Weidmann et al. 2020 and references
therein).

In 2019 we obtained high-resolution \'echelle spectra of IC\,4997, which, when compared with
the high-resolution coud\'e spectra obtained in 1993, revealed strong changes in the H$\alpha$ and
[N\,{\sc ii}] emission line profiles that had never before been reported for IC\,4997. To
follow these changes, we also obtained high-resolution \'echelle spectra in 2020. In this
paper we present the spectra in the three epochs and analyze and discuss the observed changes.

\section{Observations}

High-resolution spectra of IC\,4997 were obtained on 1993 July 16, 2019 July 31, and 2020 September 21 with the 2.2\,m
telescope at Calar Alto Observatory (Almer\'{\i}a, Spain). In 1993 we used the f/12 camera of the coud\'e spectrograph 
and a GEC 22.5\,$\mu$m CCD as detector to observe the 6540--6586\,{\AA} spectral range. The long slit, 1$\arcsec$ in width,
was centered on the object and oriented at position angles (PAs) 9$^{\circ}$, 54$^{\circ}$, 86$^{\circ}$, 279$^{\circ}$, and
324$^{\circ}$. Exposure time was 900\,s for each spectrum. Seeing was $\sim$1$\arcsec$. In 2019 and 2020 we used the \'echelle
spectrograph CAFE (Aceituno et al. 2013; Lillo-Box et al. 2020) equipped with an
IKON-L\,DZ936 detector. The 2$\farcs$4 circular
aperture of CAFE was centered on the object.
Spectra were obtained with exposure times of 60, 120, 300, and 900\,s in 2019, and 60 and 1200\,s in 2020. The H$\alpha$ emission
line is saturated in the 2019 900\,s spectrum but, due to cloudy conditions, not in the 2020 1200\,s spectrum. Seeing was
$\sim$1$\farcs$4 in 2019 and 2020. The spectral range covered by CAFE is 4070--9245\,{\AA}, although 
in this paper we concentrate on the region around the H$\alpha$ and [N\,{\sc ii}]$\lambda$$\lambda$6548,6583
emission lines for comparison with the 1993 coud\'e spectra, and on the H$\beta$ emission line in 2019 and 2020 for comparison with
the H$\alpha$ spectrum in these two epochs. A Th-Ar lamp was used for wavelength calibration in the
three epochs, and the spectra were not flux calibrated. 

The f/12 coud\'e spectra were reduced using standard procedures for long-slit spectroscopy in the {\sc iraf} package. 
The achieved spectral resolution is $\sim$12\,km\,s$^{-1}$, as indicated by the full width at half maximum (FWHM) of the Th-Ar lines in the comparison
lamp spectra, and the accuracy in radial velocity is $\pm$1\,km\,s$^{-1}$. From each long-slit coud\'e spectrum we
extracted a region of $\sim$2$\farcs$4 in size centered on the CS, and then we combined the five extracted regions into a
single spectrum for comparison with the CAFE ones. The CAFE spectra we reduced with the publicly available instrument pipeline
\texttt{cafextractor} (Lillo-Box et al. 2020). The spectral resolution (FWHM) at H$\alpha$ is $\sim$5\,km\,s$^{-1}$, and the accuracy
in radial velocity is $\sim$10\,m\,s$^{-1}$.

Throughout this paper we adopt rest wavelengths of 4861.33, 6562.82, 6548.05, and 6583.45\,{\AA } for the H$\beta$,
H$\alpha$, [N\,{\sc ii}]$\lambda$6548, and [N\,{\sc ii}]$\lambda$6583 emission lines, respectively, taken from 
the {\it Atomic Spectra Database NIST}\footnote{https://www.nist.gov/pml/atomic-spectra-database}, and all radial velocities
are quoted in the local standard of rest reference frame.

\section{Results}

Figures\,1a--1c present the spectra in the three epochs around the H$\alpha$ and [N\,{\sc ii}] emission lines,
and Fig.\,1d presents the same spectra but in a wider spectral range to show all the H$\alpha$ wings. The 1993 spectrum in
Fig.\,1d has been adapted from MTE96 because the f/12 coud\'e spectra do not cover the total extent of the H$\alpha$ wings
(see MTE96 for details). Figure\,2 shows the H$\beta$ emission line as observed in 2019, and we 
note that its profile is identical to that observed in 2020 (not shown here). Figure\,3 shows details of the [N\,{\sc ii}] emission
lines in the three epochs around the high-velocity component (HVC) previously detected by MTE96. Figures\,1 and 3 reveal noticeable
changes in the H$\alpha$ and [N\,{\sc ii}] emission line profiles in the last $\sim$27\,yr, which are described below.

The 1993 H$\alpha$ emission line presents a P\,Cygni profile with emission peaks at
$-$95.1$\pm$1.1 and $-$29.2$\pm$1.0\,km\,s$^{-1}$, the red one stronger than the blue one, separated by an 
absorption reversal at $-$79.5$\pm$1.3\,km\,s$^{-1}$ (Fig.\,1a). The profile is similar to those observed in 1990, 1991, and 1992
(Feibelman et al. 1992; Hyung et al. 1994). In 2019 and 2020, however, a single-peaked profile is observed with the intensity peak
at $-$53.1$\pm$1.1\,km\,s$^{-1}$ and a FWHM of 49.4$\pm$1.5\,km\,s$^{-1}$  (Figs.\,1b and 1c), as averaged from the 2019 and 2020 spectra.

Figure\,1 also shows that the very broad wings of the H$\alpha$ line experienced a dramatic narrowing between 1993 and 2019--2020.
This is not only recognizable at the continuum level (Fig.\,1c) but also at higher intensity levels (Figs.\,1a--1c). Moreover, the ``local continuum''
at the positions of the [N\,{\sc ii}] lines (which is traced by the H$\alpha$ wings) clearly
presents a steeper slope in 1993 than in 2019--2020 (Fig.\,3). These results show that the Rayleigh--Raman scattering was much
less efficient in 2019-2020 than in 1993.

To quantify the narrowing of the wings, we measured the full velocity width of the H$\alpha$ line at several intensity levels from
the peak intensity and at several sigma levels from the continuum in each spectrum, hereby also obtaining the centroid velocity of the H$\alpha$ line.
Table\,1 presents some values of the full velocity width in the three epochs. The errors were obtained from the noise in the profiles
at each intensity level. For the full width at zero intensity (FWZI) and full width at the 5$\sigma$ level (FW5$\sigma$) in 1993 we used the spectrum presented by MTE96, assuming that the wings are symmetric
with respect to the centroid velocity of the H$\alpha$ line, for which we obtain a value of $-$53$\pm$3\,km\,s$^{-1}$ in the three
epochs; this agrees with the velocity of the emission peak. The FWZI and FW5$\sigma$  decreased by a factor of $\sim$1.8 and $\sim$2.2
between 1993 and 2019, respectively, while the FW0.002 and FW0.0035 (see their definition in Table\,1) decreased by a 
factor of $\sim$6 and $\sim$3.6, respectively. It should be noted that the reduction in the line wings may be apparent and
due to a weakening of the Rayleigh-scattered component. In particular, taking into account that the Raman wing profile
follows ($\Delta$$\lambda$)$^{-2}$, a decrease by a factor of $\sim$2 in the line width corresponds to a reduction by a factor
of $\sim$4 in the amplitude of the wings. The velocity widths present very similar
values in 2019 and 2020 (Table\,1), although the errors are relatively large and we cannot conclude whether the narrowing
continues between 2019 and 2020.
New high-resolution spectra are necessary to study the evolution of the H$\alpha$ line. 

\begin{table}
        \centering
        \caption{Velocity width (km\,s$^{-1}$) of the H$\alpha$ emission line at different intensity levels in the three epochs.}
        \begin{tabular}{lccc} 
          \hline\hline
          Full width\tablefootmark{a} & 1993.7.16              & 2019.7.31    & 2020.9.21 \\
          \hline
          FWZI                     & 7000$\pm$300\tablefootmark{b} & 3900$\pm$350 & 3850$\pm$200 \\
          FW5$\sigma$              & 4270$\pm$160\tablefootmark{b} & 1900$\pm$170 & 1890$\pm$100 \\
          FW0.002                  & 2400$\pm$100                  & 404$\pm$30   & 400$\pm$20  \\
          FW0.0035                 & 1020$\pm$40                   & 290$\pm$17   & 282$\pm$12   \\
          FW0.005                  & 340$\pm$23                    & 235$\pm$12   & 231$\pm$10   \\
                \hline
        \end{tabular}
        \tablefoot{
        \tablefoottext{a}{FWZI=full width at the continuum level; FW5$\sigma$=full width at 5$\sigma$ level from the continuum;
        FW0.002/0.0035/0.005=full width at the 0.002/0.0035/0.005 level from the peak intensity.}
        \tablefoottext{b}{From the f/3 coud\'e spectrum assuming symmetry of the wings with respect to the radial velocity centroid of the
          H$\alpha$ emission line (see text).}
        }
\end{table}

\begin{figure}
  \centering
        \includegraphics[width=8.0cm]{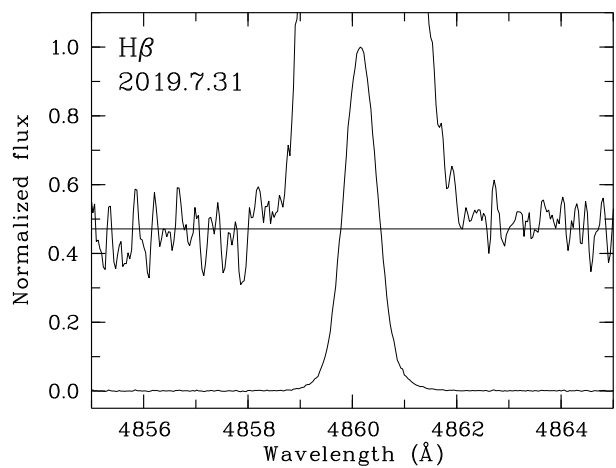}
        \caption{Normalized H$\beta$ emission line profile as observed in 2019. The profile is presented at two scales to
          show the absence of broad wings in this line. The upper profile has been smoothed, and the horizontal line represents the
          continuum obtained from emission-line-free regions at both sides of the H$\beta$ line.
}
\end{figure}

The H$\beta$ line (Fig.\,2) shows a single-peaked profile with the intensity
peak at $-$54.8$\pm$1.1\,km\,s$^{-1}$, a FWHM of 45.5$\pm$1.5\,km\,s$^{-1}$, and a FWZI of 245$\pm$25\,km\,s$^{-1}$, as averaged from the
2019 and 2020 spectra. The radial velocity and FWHM are very similar to those obtained from the H$\alpha$ line, but
the FWZI is very much smaller. Although the H$\beta$ emission is weaker and, hence, noisier than the H$\alpha$ one, there is no hint
that it could show a large FWZI. This result is consistent with Rayleigh-Raman scattering that produces wings in the H$\beta$ line very
much narrower than those in the H$\alpha$ one (e.g., Chang et al. 2018).

  \begin{figure}
  \centering
        \includegraphics[width=\columnwidth]{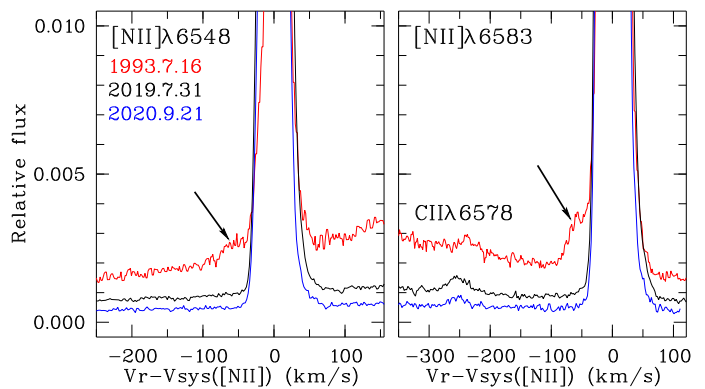}
    \caption{Details of [N\,{\sc ii}]$\lambda$$\lambda$6548,6583 emission line profiles 
      in the three epochs around the HVC observed in 1993 (arrowed). The C\,{\sc ii} emission line seems to be
        redshifted in the 1993 spectrum as compared with the 2019 and 2020 ones. The very poor S/N in the 1993 spectrum of this line
        does not allow us to state whether this shift is real. }
\end{figure}

The [N\,{\sc ii}] lines in the 1993 spectrum (Figs.\,1 and 3) show the HVC that is observed at the stellar position
in the five 2D f/12 spectra (not shown here). From a three-component Gaussian line fit to the [N\,{\sc ii}] profiles in 1993 (see below),
we obtain for the HVC a radial velocity for its intensity peak of $-$105$\pm$1\,km\,s$^{-1}$, a FWHM of 41$\pm$6\,km\,s$^{-1}$, and an intensity
ratio $I$($\lambda$6583)/$I$($\lambda$6548)$\sim$2.97 that agrees with the theoretical value of $\sim$2.95. 
Surprisingly, the HVC is missing in 2019 and 2020. This cannot be due to the different spectral resolution of the coud\'e and CAFE spectra:
the intensity of the HVC in the [N\,{\sc ii}]$\lambda$6583 ($\lambda$6548) emission line is stronger than (similar to) the intensity of
the C\,{\sc ii}$\lambda$6578 emission line in 1993, which is detected in the three epochs. If the HVC were present in 2019--2020, it
would have been detected. Most probably, the disappearance of the HVC is due to changes in its physical conditions. An increase in the
excitation between 1993 and 2019 can be ruled out because high-excitation emission lines (e.g., [O\,{\sc iii}] and [Ar\,{\sc iii}]) in
the 2019 and 2020 spectra do not show any HVC. A more plausible explanation is that the $N$$_{\rm e}$ in the HVC has decreased,
thereby decreasing the number of collisional excitations until forbidden emission lines are undetectable.

The [N\,{\sc ii}] lines present two velocity components in 2019 and 2020 (Fig.\,1). The relative intensity of the blue peak to the red peak 
is higher in 2020 ($\sim$1.2) than in 2019 ($\sim$1), perhaps due to variability. These two components contrast with the slightly asymmetric,
single-peaked profile observed in 1993 (apart from the HVC). Most probably this difference is not real but due to the different spectral
resolution of the coud\'e and CAFE spectra. In fact, a two-component Gaussian line fit (three components to the 1993 spectrum; see above)
to both [N\,{\sc ii}] lines provides identical results for the centroid (systemic) velocity, $-$48.5$\pm$1.1\,km\,s$^{-1}$
(hereafter $V$$_{\rm sys}$([N\,{\sc ii}]),
and velocity splitting, $\sim$23.5\,km\,s$^{-1}$, in the three epochs. These results agree with those obtained by MTE96. Interestingly,
the intensity peak of the H$\beta$ and H$\alpha$ lines in 2019 and 2020 is blueshifted by $\sim$5--6\,km\,s$^{-1}$ with respect to
$V$$_{\rm sys}$([N\,{\sc ii}]), and the difference is larger than the errors. If it were due to the adopted rest wavelengths, a correction of
$\sim$$-$0.1\,{\AA} would be required for those of the [N\,{\sc ii}] lines. On the other hand, the difference could be real and due to
different velocities of the regions where the Balmer and [N\,{\sc ii}] lines mainly arise.

\section{Discussion}

The H$\alpha$ P Cygni profile is indicative of a strong and
dense CS wind that produces self-absorption\footnote{MTE96 suggested that the H$\alpha$ profile consisted of two components: a blue emission peak that
was associated with the HVC components observed in [N\,{\sc ii}] because of the similar radial velocities and a red emission peak that was attributed
to the inner shell. This suggestion was based on a spectrum at lower spectral resolution than the one presented here, in which the absorption reversal
was not clearly observed. However, if the H$\alpha$ blue emission peak was related to the HVC, the H$\alpha$ line should have displayed a
single peak at $-$29\,km\,s$^{-1}$ in 2019 and 2020, which is not the case.}. This wind was active at
least from 1990 to 1993 (see Feibelman et al. 1992; Hyung et al. 1994) and
also during 1996, as indicated by the changes observed in the 3.6\,cm radio continuum emission (MT98). Moreover,
the CS wind, or a part of it, was highly
collimated, at least in 1996 (MT98). The change of a P Cygni profile to a single-peaked profile indicates that the wind has dramatically weakened
and is unable to produce self-absorption. Nevertheless, a detailed interpretation of the H$\alpha$ profile is difficult because at least three
  components may be contributing: the outer and inner shell and the stellar wind. Spatially resolved, high-resolution spectra could disentangle the
  different contributions to the H$\alpha$ emission profile (see, e.g., Solf 2000).

The strength of the CS wind is related to the value of the [O\,{\sc iii}]$\lambda$4363/H$\gamma$ ratio (A+20, their Fig.\,4): 
strong wind is observed around 1990--1996 when that ratio reached its highest values, while weak wind is
observed in 2019--2020 when that ratio reached its minimum values. This result strongly favors the
variability of IC\,4997 being mainly due to a variable CS wind and the corresponding induced changes in $T$$_{\rm e}$ and $N$$_{\rm e}$ (see also A+20).
Noticeably, the C\,{\sc iv}\,$\lambda$\,1550\,{\AA} emission line did not show a P\,Cygni profile in 1978--1981, when the [O\,{\sc iii}]/H$\gamma$
ratio was  close to its minimum (Feibelman 1982, A+20), indicating that no strong CS wind was present in those years.
Nor did the C\,{\sc iv} line show a P\,Cygni profile in 1987, when the  [O\,{\sc iii}]/H$\gamma$ ratio was close to maximum, though the
N\,{\sc v}$\lambda$1238\,{\AA} emission line may have shown a P\,Cyg profile in 1981, when no strong wind should have been expected
(Marcolino et al. 2007). These results may indicate that the strength of the CS wind presents small fluctuations, as suggested by the small changes in
the H$\alpha$ emission profile (Hyung et al. 1994). These small wind fluctuations may contribute to the small fluctuations in the
[O\,{\sc iii}]$\lambda$4363/H$\gamma$ ratio observed within a year, together with other phenomena, such as nebular expansion, some changes in $T$$_{\rm eff}$, and/or
interaction between the shells (Ferland 1979; MTE96, MT98; A+20 and references therein). 

According to LH00, the broad H$\alpha$ wings observed in 1991 require a column density of $\sim$2$\times$10$^{20}$\,cm$^{-2}$ in the H\,{\sc i}
envelope around IC\,4997 and an incident Ly$\beta$ flux $\ga$10$^{35}$\,erg\,cm$^{-2}$\,s$^{-1}$ that should be 
generated in a high electron density ($\ga$10$^{9}$\,cm$^{-3}$), compact ionized region located very close ($\sim$0.1\,AU) to the CS.
Changes in the H\,{\sc i} column density and/or Ly$\beta$ flux would then be responsible for the narrowing of the H$\alpha$ wings. The column density would
be reduced with time due to the expansion of the H\,{\sc i} envelope and to the photoionization of its innermost region by the advance of the ionization
front. The H\,{\sc i} envelope expands at $\sim$14\,km\,s$^{-1}$ with respect to $V$$_{\rm sys}$([N\,{\sc ii}]) (Altschuler et al. 1986), and
in 26\,yr the increase in radius would be $\sim$3.7$\times$10$^{-4}$\,pc. If we assume a mean nebular radius of 1$\farcs$2 ($\sim$1.5$\times$10$^{-2}$\,pc
at a distance of 2.5\,kpc) as the minimum size of the H\,{\sc i}, and that the column density varies as 1/R$^2$, the decrease in
the column density in 26\,yr would be $\sim$1.05. In addition, if the ionization front expands with the same velocity as the outer
shell ($\sim$12\,km\,s$^{-1}$ from the [N\,{\sc ii}] lines; MTE96), the increase in the nebular radius would be $\sim$3.2$\times$10$^{-4}$\,pc, which would
imply a reduction in the column density by a factor  of $\sim$1.04 in 26\,yr. By combining the two processes, we can estimate that the column density would have decreased by a
factor of $\sim$1.1 over the last 26 yr. Although this number is an approximation, values of this order cannot account for the observed
decrease in the line wings, which requires a much larger factor (see LH00, their Fig.\,1). Therefore, the electron density in the very compact
and dense region around the CS should have decreased by, for example, the expansion of that region,
reducing the Ly$\beta$ flux and hence the efficiency of the
Rayleigh--Raman scattering.
  
An interesting question regards the origin of a variable CS wind in IC\,4997. The variation in the
[O\,{\sc iii}]$\lambda$4363/H$\gamma$ ratio suggests
that, in addition to its maximum value around 1995, another maximum may have
existed around 1938--1940 when that ratio reached values very similar to
those in 1995 (Feibelman et al. 1979; A+20, their Fig.\,4). The same can be said for
the minimum values of the ratio, around 1965 and 2020, although more points in the coming years are necessary
to confirm periodicity. As of now, the observed variability of that ratio may be better considered as episodic. 
  Moreover, despite the presence of small fluctuations, the [O\,{\sc iii}]$\lambda$4363/H$\gamma$ ratio 
  has shown a general continuous and relatively smooth variation between its extreme values. The same is observed in
  the [O\,{\sc iii}]$\lambda$4363/H$\beta$, 
  [O\,{\sc iii}]$\lambda$4959/H$\beta$, and [O\,{\sc iii}]$\lambda$4363/[O\,{\sc iii}]$\lambda$4959 line intensity ratios,
  as well as in the intensity of
  the [O\,{\sc iii}]$\lambda$$\lambda$4363,4959 and H$\beta$ emission lines. These variations are less compatible with a sudden outburst but
  indicate that the CS wind should vary in an episodic and smooth manner too. The variability in the emission lines and emission line ratios
  points to a binary CS with a period of $\sim$50--60\,yr. The idea of a binary (or triple) CS in IC\,4997 has been suggested
  before (MT98; Bear \& Soker 2017;
  A+20 and references therein). The results presented in this paper may help to refine the binary scenario, if we assume a
  companion in an eccentric orbit
  with a period of $\sim$50--60\,yr. Around 1940 and 1995 the stars were at their minimum separation; the CS wind may have been
  activated by the close presence of the companion, and, in addition, mass transfer from the companion could have formed (or fed)
  a kind of very dense, ionized
  accretion disk-like structure around the CS, which generated enough Ly$\beta$ photons to produce the large
  H$\alpha$ wings. This dense region could, at least partially,
  collimate the CS wind, and we note that, in this scenario, the collimating agent would
  be related to the CS and not to the companion. Ejection of high-velocity knots from the dense disk-like structure may account for the HVC
  observed in 1993. MT98 and G\'omez et al.
  (2002) suggested that the HVC could be associated with a faint nebular clump observed in the radio continuum maps at 7\,mm and 2\,cm, with
  a low electron density of $\sim$700\,cm$^{-3}$ (MTE96). If the size of this clump increases with time, its $N$$_{\rm e}$ will decrease
  until emission lines
  are no longer detected after a certain time. As the distance between the stars increases, the influence of the companion on
  the CS and the mass transfer would decrease until reaching a minimum around 1965 and 2020, at the maximum binary separation. As a consequence,
  the CS wind would be weak, and the very dense ring-like structure around the CS would no longer be fed and may expand, hence decreasing its
  electron density and the Ly$\beta$ flux and resulting in a low efficiency of the Rayleigh--Raman scattering. 

If the scenario described above is correct, and the physical conditions 
approximately repeat themselves every 50--60\,yr, an increase in the [O\,{\sc iii}]$\lambda$4363/H$\gamma$ ratio will be
detectable in $\sim$5\,yr, and the
expected strengthening CS wind might manifest itself as asymmetries in the H$\alpha$ line profile and broader wings.
Monitoring of the [O\,{\sc iii}]$\lambda$4363/H$\gamma$ ratio (and more emission lines) and of the H$\alpha$ line
profile in the coming years is crucial to
confirming this scenario. Furthermore, many multiwavelength observations of IC\,4997 exist around 1995 when the CS wind was strong. Similar
multiwavelength observations carried out now, when the CS wind is weak, will provide key information about the ongoing processes in IC\,4997
and their possible relationship to a long period binary CS.

\section{Conclusions}

High-resolution spectra of the variable PN IC\,4997 obtained in 1993, 2019, and 2020 have revealed that between 1993 and 2019--2020: (1) the
H$\alpha$ P\,Cygni emission line profile observed in 1993 changed to a single-peaked emission line profile, indicating that the stellar wind has dramatically
weakened in the last 26\,yr; (2) the broad wings of the H$\alpha$ emission lines narrowed by a factor of $\sim$2, showing that the Rayleigh--Raman scattering
process was much less efficient in 2019--2020 than in 1993, most probably due to a decrease in the Ly$\beta$ flux necessary to produce large line wings; and (3) a
high-velocity nebular component detected in 1993 in the [N\,{\sc ii}] emission lines is missing in 2019--2020, which may be attributed to a large decrease
in its electron density. We found a relationship between the strength of the CS wind and the extreme values of the episodically variable
[O\,{\sc iii}]$\lambda$4363/H$\gamma$ ratio, other line intensity ratios, and emission line intensities, which strongly suggests that an episodic, smoothly variable
wind is the main cause of the variability of IC\,4997. A companion in an eccentric orbit would account for many of the observed characteristics of IC\,4997,
at least qualitatively. Monitoring of the [O\,{\sc iii}]$\lambda$4363/H$\gamma$ ratio and the H$\alpha$ emission profile in the coming years is
crucial to confirming this scenario. Moreover, important information about the ongoing processes in IC\,4997 will be reached by comparing
multiwavelength observations carried out now, when the CS wind is weak, with similar observations obtained $\sim$25--30\,yr ago, when the CS wind was strong. 

\begin{acknowledgements}
 We are very grateful to our referee, William Henney, for his prompt answer and useful comments that have improved the paper.
  We thank Calar Alto Observatory for allocation of director's discretionary
time to this programme. LFM is partially supported by MCIU grant AYA2017-84390-C2-1-R (co-funded by FEDER) and PID2020-114461GB-I00 of AEI
(10.13039/501100011033), and acknowledges financial support from the State Agency for Research of the Spanish MCIU through the "Center of Excellence
Severo Ochoa" award for the Instituto de Astrof\'{\i}sica de Andaluc\'{\i}a (SEV-2017-0709). JMT acknowledges partial support from the PID2020-117710GB-I00
grant funded by MCIN/ AEI /10.13039/501100011033. J.L-B. acknowledges financial support received from "la Caixa" Foundation (ID 100010434) and from the
European Unions Horizon 2020 research and innovation programme under the Marie Slodowska-Curie grant agreement No 847648, with fellowship
code LCF/BQ/PI20/11760023. This research has also been partly funded by the Spanish State Research Agency (AEI) Projects No.ESP2017-87676-C5-1-R and
No. MDM-2017-0737 Unidad de Excelencia "Mar\'{\i}a de Maeztu"- Centro de Astrobiolog\'{\i}a (INTA-CSIC).
\end{acknowledgements}

%
%

\end{document}